\documentclass[12pt]{lncse} 
\usepackage{multicol} 
\usepackage{makeidx}  
\usepackage{amsmath}
\usepackage{graphicx}
\usepackage{amssymb}
\usepackage{latexsym}
\makeindex
\renewcommand{\vec}[1]{\relax\ifmmode\mathchoice
{\mbox{\boldmath$\relax\displaystyle#1$}}
{\mbox{\boldmath$\relax\textstyle#1$}}
{\mbox{\boldmath$\relax\scriptstyle#1$}}
{\mbox{\boldmath$\relax\scriptscriptstyle#1$}}\else
\hbox{\boldmath$\relax\textstyle#1$}\fi}
\newcommand{\be}{\begin{equation}}
\newcommand{\ee}{\end{equation}}
\newcommand{\bea}{\begin{eqnarray}}
\newcommand{\eea}{\end{eqnarray}}
\newcommand{\bdm}{\begin{displaymath}}
\newcommand{\edm}{\end{displaymath}}

\newcommand{\bi}{\begin{itemize}}
\newcommand{\ei}{\end{itemize}}

\newcommand{\refkl}[1]{(\ref{#1})}

\newcommand{\sub}[1]{_{\rm #1}}
\renewcommand{\sup}[1]{^{\rm #1}}

\begin{document}
\unitlength 0.9mm
\title{An adaptive smoothing method for traffic state identification 
from incomplete information}
\titlerunning{Adaptive smoothing method for traffic state identification}
\author{Martin Treiber \and Dirk Helbing}
\institute{Institute for Economics and Traffic,
Faculty of Traffic Sciences ``Friedrich List'',\\
Dresden University of Technology, D-01062 Dresden, Germany}
\maketitle
\begin{abstract}
We present a new method to obtain spatio-temporal information
from aggregated data of stationary traffic detectors, the 
{\em ``adaptive smoothing method''.}
In essential, a nonlinear spatio-temporal lowpass filter
is applied to the input detector data. This filter exploits the fact that, in
congested traffic, perturbations travel upstream at a constant
speed, while in free traffic, information propagates downstream. 
As a result, one obtains velocity, flow,
or other traffic variables as smooth functions of space and time.
Applications include traffic-state visualization, 
reconstruction  of traffic situations 
from incomplete information, fast identification of traffic
breakdowns (e.g., in incident detection), 
and experimental verification of traffic models.

We apply the adaptive smoothing method to observed congestion patterns 
on several German freeways.
It manages to make sense out of data where conventional
visualization techniques fail.
By ignoring  up to \mbox{65\%} of the detectors
and applying the method to the reduced data set,
we show that the results are robust. The 
method works well if the distances between neighbouring
detector cross sections do not exceed 3 km.
\end{abstract}

\section{Introduction}
During the last decades, traffic dynamics and pedestrian flows have been intensively
studied. Regarding the observed phenomena and simulation approaches we refer the
reader to some recent reviews \cite{Helb-Opus,Schad,PED,ENV}.
Presently, scientists are more and more getting interested in 
detailed empirical studies. Reasons for this are the
availability of better data and the need to verify and calibrate models.
Both, theoretical \cite{Phase,Lee3LeKi99,PINCH} and empirical 
\cite{CasBe99,DagCaBe99,Hal1Ag91,KernerPinch,Kerner-Rehb96a,Kerner-Rehb96b,Koshi,Lee3LeKi00,%
NeubSaScSc99,Opus}
studies indicate that the phenomenology of congested traffic is more complex
than originally expected. There seems to be 
a rich spectrum of traffic states \cite{Phase,Lee3LeKi99,Kerner-Rehb96a,Kerner-Rehb96b},
hysteretic or continuous temporal or spatial transitions among them 
\cite{PINCH,KernerPinch,Kerner-Rehb96a,Kerner-Rehb96b,HelTr98a,Lee3LeKi98}, 
and fluctuations or a erratically appearing dynamics play a significant role 
\cite{Kerner-Rehb96b,NeubSaScSc99,Bank99,Leu88,TreHe99a}.
In order to make sense out of this, it is increasingly important to have suitable
ways of data processing, in order to extract the information relevant for scientific
investigations or specific applications. Here, we will propose a 
three-dimensional
data-evaluation method allowing to visualize the spatio-temporal dynamics of
traffic patterns along freeways. 
\par
The developed {\em ``adaptive smoothing method''} filters out small-scale fluctuations
and adaptively takes into account the main propagation direction of the information flow
(i.e. the dominating characteristic line), which have been determined by means of 
a spatio-temporal correlation analysis in other studies \cite{Sollacher-Status99}. The temporal 
length scale of the smoothing procedure can be as small as the sampling interval, while
the spatial length scale is related to the distance between successive detectors, which can
be up to 3 kilometers long. 
\par
By ``filter'' we just mean a transformation of the data with specific properties.
Here, we use a spatio-temporal lowpass filter, i.e., only (Fourier) components of low frequency
can pass the filter, while high-frequency contributions are considered as fluctuations and
smoothed out. One particular feature of our filter is that it is nonlinear and
adaptive to the traffic situation in distinguishing free and congested traffic, as the
propagation direction of perturbations differs. The results are three-dimensional
visualizations of traffic patterns, which are quite robust with respect to variations of
the filter parameters and very helpful in obtaining a clear picture
of the systematic spatio-temporal dynamics.
\par
Therefore, this method is suitable for the reconstruction  of traffic situations 
from incomplete information, fast identification of traffic
breakdowns (incident detection), 
and experimental verification of traffic models. First results support the
phase diagram of traffic states occuring at bottlenecks \cite{Opus,Synchro}, which,
apart from free traffic, predicts pinned or moving localized clusters, spatially extended
patterns such as triggered stop-and-go waves, and oscillating or homogeneous congested traffic,
or a spatial coexistence of some of these states \cite{Phase,PINCH,Helb-Opus}. 
The respectively occuring spatio-temporal pattern depends on the specific
freeway flow and bottleneck strength, 
but also on the level of fluctuations, as these
can trigger transitions from, for example, free traffic to localized cluster states
\cite{Phase,Opus,Helb-Opus}. We do not see sufficient support for one ``generalized pattern'' 
\cite{KerMic} that would always be observed when traffic flow breaks down. 

\section{\label{sec_method} Description of the Method}

The adaptive smoothing method is a data processing method
for obtaining traffic variables as smooth functions of space and time
out of stationary traffic data. It has
following heuristically motivated properties:

\begin{enumerate}
\item In case of free traffic, perturbations 
(of, e.g.,  velocity or flow)
move essentially into the direction of traffic flow \cite{Dag99a}.
More specifically, they propagate with a characteristic velocity $c\sub{free}$
at about 80\% of the {\it desired} velocity $V_0$ on empty roads
\cite{Sollacher-Status99}. 
Therefore, at locations with free traffic,
perturbations with propagation velocities near $c\sub{free}$
should pass the filter.
\item In case of congested traffic, perturbations propagate against the
  direction of traffic flow with a characteristic and remarkably constant
  velocity $c\sub{cong}\approx -15$ km \cite{Kerner-Rehb96a}.
With modern data analysis techniques, it has been shown that such 
propagation patterns persist even in ``synchronized'' congested traffic flow,
where they are hardly visible
in the time series due to a wide scattering of the data in this state
\cite{Dag99a}.
So, for high traffic densities or low
  velocities, the filter should transmit spatio-temporal
perturbations propagating with velocities near $c\sub{cong}$
more or less unchanged.
\item The filter should
 smooth out all high-frequency fluctuations in $t$
  on a  time scale smaller than $\tau$ and 
spatial fluctuations in
 $x$ on a length scale smaller than
  $\sigma$. The parameters $\tau$ and $\sigma$ of the smoothing method
can be freely chosen in a wide range (cf. Table \ref{tab:param}).
\end{enumerate}


Let us assume that aggregated detector data 
$\vec{z}\sup{in}_{ij}$ are available from $n$ cross sections $i$
at positions $x_i$
where $j\in\{j\sub{min}, \cdots, j\sub{max}\}$ 
denotes the index of the aggregation
intervals.

Usually, the  aggregation interval
\begin{equation} 
\Delta t = t_j - t_{j-1}
\end{equation}
is fixed  (between 20 s and 5 min, depending on the measurement device). 
A typical value for German highways is $\Delta t=1$ min.

The components of $\vec{z}\sup{in}_{ij}$ 
represent the desired aggregated quantities as
obtained from cross section $i$
during the time interval $j$. Typical examples include the average
velocity $V_{ij}$, the vehicle flow $Q_{ij}$, the occupancy $O_{ij}$, or some
derived quantity such as the traffic density $\rho_{ij}$. The input data
 $\vec{z}\sup{in}_{ij}$
can represent either averages over all lanes or quantities on a given lane.

The adaptive smoothing  method provides estimates $\vec{z}(x,t)$ for
{\it all} locations $x\in [x_1, x_{n}]$ between the positions of
the first and the last detector, and for all times 
$t\in [t\sub{min},t\sub{max}]$.
Extrapolations 
(in the sense of a short-term traffic forecast) 
will also be discussed.
Without loss of generality, we assume $x_1<x_2<\cdots <x_{n}$,
and a traffic flow in positive $x$-direction. 


\vspace{5mm}

The core of our {\em ``adaptive smoothing method''} is a
nonlinear filter transforming the discrete input detector data
$\vec{z}_{ij}\sup{in}$ 
into the smooth spatio-temporal functions $\vec{z}(x,t)$.
To satisfy the first two requirements mentioned above, 
we write the  filter as
\begin{equation}
\label{zgen}
\vec{z}(x,t) =  w(\vec{z}\sub{cong},\vec{z}\sub{free}) 
                \vec{z}\sub{cong} (x,t)
             + \big[1-w(\vec{z}\sub{cong},\vec{z}\sub{free}) \big]
                \vec{z}\sub{free} (x,t).
\end{equation}
This  is a superposition of two linear anisotropic lowpass filters
$\vec{z}\sub{cong} (x,t)$ and $\vec{z}\sub{free} (x,t)$ 
with an adaptive weight factor $0\le w\le 1$ which itself depends
nonlinearly on the output of the linear filters as discussed later on.

The filter $\vec{z}\sub{cong} (x,t)$  for congested traffic is given by
\begin{equation}
\label{zcong}
\vec{z}\sub{cong}(x,t) = \frac{1}{N\sub{cong}(x,t)} 
  \sum\limits_{i=1}^{n} 
  \sum\limits_{j=j\sub{min}}^{j\sub{max}}
\phi\sub{cong}(x_i - x,  t_{j} - t)
\vec{z}\sup{in}_{ij},
\end{equation}

with
 
\begin{equation}
t_j=t\sub{min}+ j \Delta t,
\end{equation}

and the normalization factor

\begin{equation}
\label{Ncong}
N\sub{cong}(x,t) =
  \sum\limits_{i=1}^{n} 
  \sum\limits_{j=j\sub{min}}^{j\sub{max}}
\phi\sub{cong}(x_i - x,  t_{j} - t).
\end{equation}
This normalization guarantees that a constant input vector 
$\vec{z}\sup{in}_{ij}=\vec{z}_0$ for all $ i, j$
 is transformed into a constant output function with the same components:
$\vec{z}(x,t)=\vec{z}_0$.
Notice that the normalization depends on both, location $x$ and time $t$.

The analogous expression for the lowpass filter $\vec{z}\sub{free} (x,t)$
for free traffic is
\begin{equation}
\label{zfree}
\vec{z}\sub{free}(x,t) = \frac{1}{N\sub{free}(x,t)} 
  \sum\limits_{i=1}^{n} 
  \sum\limits_{j=j\sub{min}}^{j\sub{max}}
\phi\sub{free}(x_i - x,  t_{j} - t)
\vec{z}\sup{in}_{ij}
\end{equation}
with
\begin{equation}
\label{Nfree}
N\sub{free}(x,t) =
  \sum\limits_{i=1}^{n} 
  \sum\limits_{j=j\sub{min}}^{j\sub{max}}
\phi\sub{free}(x_i - x,  t_{j} - t).
\end{equation}

\begin{figure}[htbp] 
\begin{center}
    \includegraphics[width=120\unitlength]{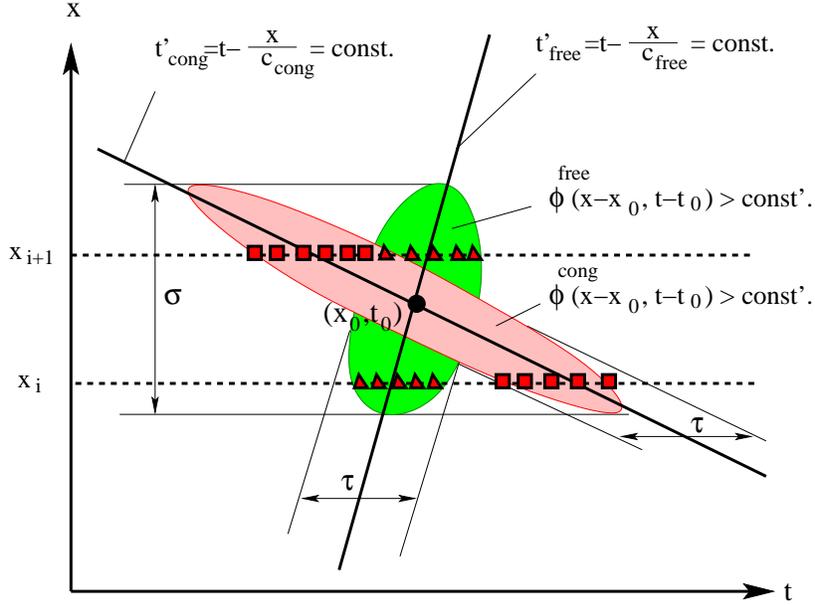} 
\end{center}
\caption[]{\label{filterSketch}Visualization of the effects of
linear homogeneous filters with the kernels
$\phi\sub{free}(x,t)$ and $\phi\sub{cong}(x,t)$, respectively.
The shaded areas denote the regions considered in the calculation
of a data point 
at $(x,t)$. Triangles denote the mainly contributing 
input data sampled in  free  traffic, squares the ones sampled
in congested traffic.}
\end{figure}

The kernels  $\phi\sub{cong}(x,t)$ and 
$\phi\sub{free}(x,t)$ of the
linear homogeneous filters do the required smoothing and
are particularly transmissible for  perturbations
propagating with the typical velocities 
$c\sub{cong}$ and $c\sub{free}$ observed in congested and free
traffic, respectively
(cf. Fig. \ref{filterSketch}). 

If the propagation velocity of the perturbations was zero, the filter should
only perform the smoothing. We chose the (non-normalized)
exponential function
\begin{equation}
\label{phi0}
\phi_0(x,t) = \exp\left(-\frac{|x|}{\sigma}
                        -\frac{|t|}{\tau}\right).
\end{equation}
Instead, one could apply other localized functions
such as a two-dimensional  Gaussian. However, it  turned out that the
exponential had more favourable properties in our application.

The linear filter for nonzero propagation velocities can be mapped to
$\phi_0(x,t)$ by the coordinate transformations 
(cf. Fig. \ref{filterSketch})
\begin{equation}
\label{transform}
x=x', \ \ \ t = t'_{\rm cong} + \frac{x}{c\sub{cong}}
                     = t'_{\rm free} + \frac{x}{c\sub{free}}.
\end{equation}
Thus, we obtain for the kernels of the linear anisotropic filters
\begin{eqnarray}
\label{phicong}
\phi\sub{cong} (x,t) &=& \phi_0(x', t'_{\rm cong})
= \phi_0\left(x,t - \frac{x}{c\sub{cong}}\right), \\
\label{phifree}
\phi\sub{free} (x,t) &=& \phi_0(x', t'_{\rm free})
= \phi_0\left(x,t - \frac{x}{c\sub{free}}\right).
\end{eqnarray}

\begin{figure}[htbp] 
\begin{center}
    \includegraphics[width=100\unitlength]{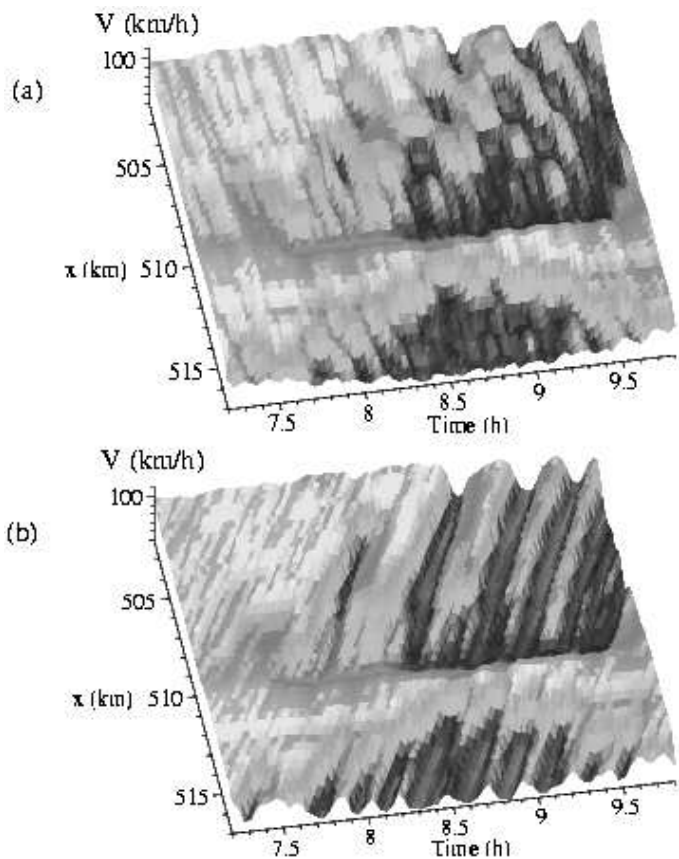} \\[0mm]
\end{center}
\caption[]{\label{vFreevCong}Typical velocity fields
$V\sub{free}(x,t)$ (top) and  $V\sub{cong}(x,t)$ (bottom)
obtained by application of the filters $\vec{z}\sub{free} (x,t)$ and 
$\vec{z}\sub{cong} (x,t)$ to traffic data of a section of the
German freeway A9 South.
}
\end{figure}

Figure \ref{vFreevCong} shows the action of the filters
$\vec{z}\sub{cong} (x,t)$ and $\vec{z}\sub{free} (x,t)$ 
for the velocity fields
$V\sub{cong} (x,t)$ and $V\sub{free} (x,t)$.

Finally, we define the nonlinear adaptive weight function
$w(\vec{z}\sub{cong}, \vec{z}\sub{free}) \in [0,1]$.
Obviously, we must have
$w\approx 1$ for congested traffic, and
$w\approx 0$ for free traffic, so we need some {\it a priori}
estimate of the traffic situation at the point $(x,t)$.

Congested traffic is characterized by a high traffic density and low
average velocity. Since, in contrast to the density, 
the velocity can be directly measured with stationary detectors, we 
chose the velocity to determine the {\it a priori} estimate.
Different possibilities to estimate the velocity at point $(x,t)$ are:
\bi
\item
The measured velocity $V_{ij}$ of the detector cross section
 whose position $x_i$ is nearest to $x$  in the time interval $j$
containing the actual time $t$,
\item the velocity  $V\sub{cong}(x,t)$ as calculated with the
``congested-traffic'' filter $\vec{z}\sub{cong}$ according to Eq.~\refkl{zcong} 
 with kernel \refkl{phicong}
  (assuming {\it a priori} congested traffic),
\item the velocity $V\sub{free}(x,t)$ as calculated with the
``free-traffic'' filter $\vec{z}\sub{free}$ according to Eq.~\refkl{zfree} 
 with the kernel \refkl{phifree},
\item or some combination of the above estimates.
\ei
The first way to estimate the velocity is subject to errors,
if the typical length scale 
$\lambda$ of occuring stop-and-go structures is not larger than
the distance $\Delta x_i$ between two neighbouring detectors.
At this point,  it is crucial  that propagating structures in 
congested traffic, especially stop-and-go waves, are very persistent.
It has been shown \cite{Ker00} that they can propagate 
through freeway intersections or other inhomogeneities nearly unchanged,
passing all perturbations of free traffic on their way (see, e.g., 
Fig. \ref{A5North}).
Therefore, whenever at least one of the estimates indicates
congested traffic, the weight function should favour the
filter for congested traffic. Specifically, we assume
\begin{equation}
\label{w}
w(\vec{z}\sub{cong},\vec{z}\sub{free}) 
  = w(V\sub{cong}(x,t),V\sub{free}(x,t)) 
  = \frac{1}{2}\left[
   1+\tanh\left(\frac{V_c - V^*(x,t)}{\Delta V} \right)\right] \, ,
\end{equation}
where
\begin{equation}
\label{vDecide}
V^*(x,t) = \min \bigg(V\sub{cong}(x,t),V\sub{free}(x,t) \bigg) \, .
\end{equation}
$V_c$ and $\Delta V$ are parameters that can be varied in a wide
range.

If not explicitely stated otherwise, for all 
simulations in the following section we will use the
parameters specified in Table \ref{tab:param}.
We will also show that the four parameters $c\sub{cong}$,
$c\sub{free}$, $V_c$ and $\Delta v$ can be
varied in a wide range without great differences in the output.
In this way, we show that the proposed adaptive smoothing method
does not need to be calibrated to the respective freeway. One can take the
values from Table \ref{tab:param} as a global setting.

The smoothing parameters $\sigma$ and $\tau$ have the same meaning and
the same effect as in standard smoothing
methods, e.g., Eqs. (22) and (23) in
Ref. \cite{Opus}.


\vspace{0mm}
\begin{table}

\newcommand{\entry}[3]{
                        \parbox{25mm}{\vspace*{2mm}#1\vspace*{2mm}}
                      & \parbox{25mm}{\vspace*{2mm}#2\vspace*{2mm}}
                      & \parbox{70mm}{\vspace*{2mm}#3\vspace*{2mm}} 
                        \\ \hline
                       }
\begin{center}
\begin{tabular}{|l|l|l|} 
\hline
\entry{Parameter}{Typical Value}{Meaning}
\hline
\hline
\entry{$\sigma$}{0.6 km}{Range of spatial smoothing in $x$} 
\entry{$\tau$}{1.1 min}{Range of temporal smoothing in $t$} 
\entry{$c\sub{free}$}{80 km/h}
      {Propagation velocity of perturbations in free traffic}
\entry{$c\sub{cong}$}{$-15$ km/h}
      {Propagation velocity of perturbations in congested traffic}
\entry{$V_c$}{60 km/h}{Crossover from free to congested traffic}
\entry{$\Delta V$}{20 km/h}{Width of the transition region}
\end{tabular}
\end{center}
\caption[]{\label{tab:param}Parameters of the {\em ``adaptive smoothing method''} defined by
Eqs.~\protect\refkl{zgen}--\protect\refkl{vDecide}, their
interpretation, and their typical values.
}
\end{table}
In summary, the proposed {\em ``adaptive smoothing method''} is given by
Eq. \refkl{zgen} with the nonlinear weight function \refkl{w}, 
the filter \refkl{zcong} with normalization \refkl{Ncong} and kernel
\refkl{phicong} for congested traffic, the filter \refkl{zfree} 
with normalization \refkl{Nfree} and kernel \refkl{phifree} for free
traffic,  and the smoothing filter \refkl{phi0}.
Table \ref{tab:param} gives an overview of the six parameters involved
and typical values for them.
The adaptive smoothing method includes the following special cases:
\bi
\item Isotropic smoothing resulting in the limits $c\sub{free}\to\infty$ and
$c\sub{cong}\to\infty$ (for practical purposes, one may chose
$c\sub{free}=c\sub{cong}=10^6$ km/h);
\item only filtering for structures of congested traffic in the limit 
$V_c\to\infty$ (for practical purposes, one may set $V_c\gg V_0$ 
with $V_0$ being the desired velocity);
\item only filtering for structures of free traffic for 
$V_c=\Delta V=0$;  
\item consideration of the data of the nearest detector only, if
$\sigma=0$;
\item application of the actual sampling interval of the detectors,
if $\tau=0$, $c\sub{free}\to\infty$ and
$c\sub{cong}\to\infty$.
(To avoid divisions by zero, zero values of $\sigma$, $\tau$, and
$\Delta V$ are replaced by very small positive values in the software).
\ei

\section{\label{sec_appl}Application to German Freeways}

We will now discuss data from the German freeways A8-East and A9-South near Munich,
and of the freeway A5-North near Frankfurt.
In all cases, the traffic data were obtained
from several sets of 
double-induction-loop detectors recording, separately for each lane,
the passage times and velocities of all vehicles. 
Only aggregate information was stored with an aggregation interval
of $\Delta t=1$ min.
We will use the following input data $\vec{z}_{ij}\sup{in}$:
\bi
\item The lane-averaged vehicle flow
\begin{equation}
\label{Q_emp}
Q_{ij} = \sum_{l=1}^L \frac{n_{ij}^{l}}{L\Delta t} \, ,
\end{equation}
where $n_{ij}^{l}$ is the vehicle count at cross section $i$ during time
interval $j$ on lane $l$. The considered sections of the
freeways have $L=3$  lanes in most cases.
\item
The lane-averaged mean velocity
\begin{equation}
\label{v_emp}
V_{ij} = \sum_{l=1}^L \frac{Q_{ij}^{l} V_{ij}^{l}}{Q_{ij}} \, ,
\end{equation}
where $V_{ij}^{l}$ is the average velocity at cross section $i$ during time
interval $j$ on lane $l$. 
\item
The traffic density determined via the formula
\begin{equation}
\label{rho_emp}
\rho_{ij} = \frac{Q_{ij}}{V_{ij}} \, ,
\end{equation}
as occupancies were not available for all freeways.
\ei


\begin{figure}[htbp] 
\begin{center}
\includegraphics[width=100\unitlength]{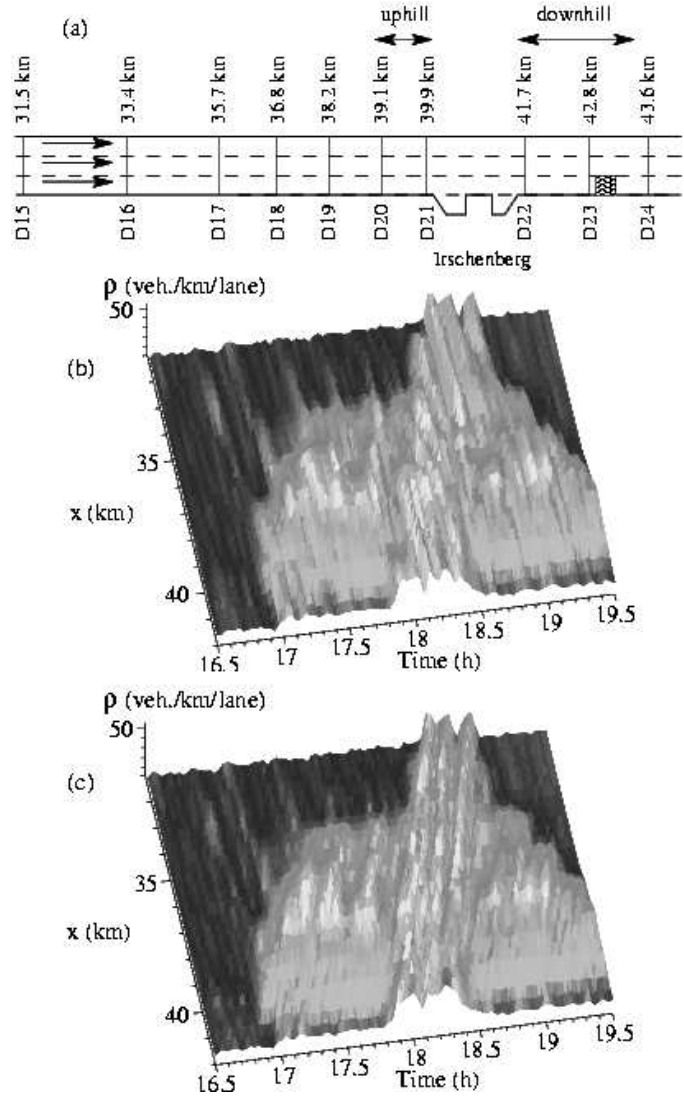}
 \end{center}
\caption[]{\label{flugzeug}Complex  congested traffic pattern on the 
German freeway A8-East
from Munich to Salzburg during the evening rush hour
{on} November 2, 1998.
(a) Sketch of the freeway. 
(b) Plot of the spatio-temporal density $\rho(x,t)$ using conventional
smoothing (resulting for the setting $c\sub{free}=c\sub{cong}=10^6$ km/h).
(c) Plot of the  same data 
as calculated with the adaptive smoothing method.
}
\end{figure}

Figure \ref{flugzeug} shows an example of a  complex
traffic breakdown that occurred on the
freeway A8 East from Munich to Salzburg during the evening rush hour
{on} November 2, 1998. 
Two different kinds of bottlenecks were relevant, (i) a relatively
steep uphill gradient from $x=38$ km to $x=40$ km
(the ``Irschenberg''), and (ii) an
incident leading to the closing of one of the three lanes between
the cross sections D23 and D24 between $t=$\,17:40 h and $t=$\,18:10 h.
For further details, see Ref.~\cite{Opus}.
Figure \ref{flugzeug}(c) shows the traffic density using the
proposed data processing method with the parameters specified in
Table~\ref{tab:param}. For comparison, 
(b) shows the result of conventional isotropic smoothing ignoring the
speed of information propagation, cf.
Ref.~\cite{Opus}.
While the structures in the free-traffic regions are nearly the same
in both plots, the new smoothing method  gives a better reconstruction of the
 congested traffic patterns.
In contrast to the conventional method, 
our proposed method can resolve 
oscillations with wavelengths $\lambda$ comparable to or smaller
than the distance $\Delta x_i$ between two neighbouring cross
sections as illustrated  for the three
stop-and-go waves propagating through the whole 
displayed section ($\lambda \approx 2$~km).
Notice that the conventional method generates  artefacts for this 
example. 

The congestion patterns caused by the uphill gradient (the ``wings'' in
Fig. \ref{flugzeug}(b), (c)) have larger characteristic wavelengths 
and are resolved, at least partially, also with the conventional method.

\begin{figure}[htbp] 
\begin{center}
    \includegraphics[width=100\unitlength]{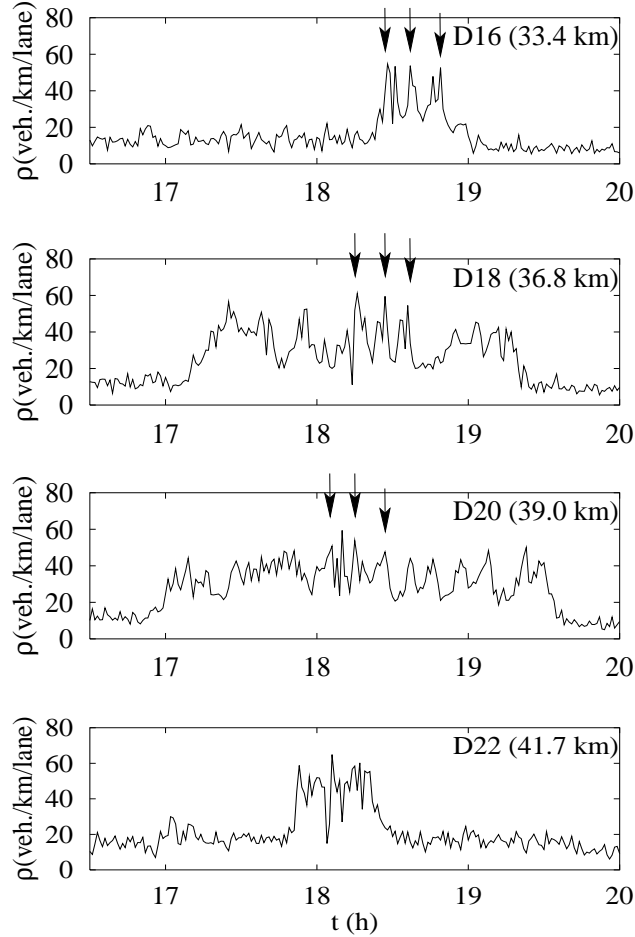}
\end{center}
\caption[]{\label{flugzeug_t}Time series of the empirical density,
approximated by means of Eq. \protect\refkl{rho_emp},  for several cross sections.}
\end{figure}

\begin{figure}[htbp] 
\begin{center}
    \includegraphics[width=100\unitlength]{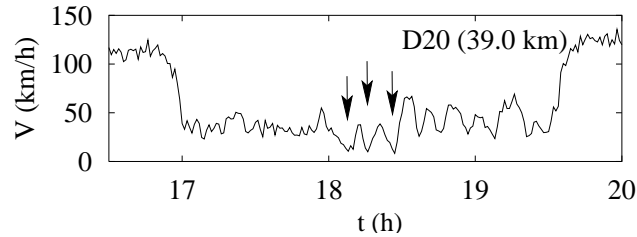}
\end{center}
\caption[]{\label{flugzeugV_t}Time series of the empirical velocity data
determined via Eq. \protect\refkl{v_emp} for cross section D20.
}
\end{figure}

For comparison, we plot time series of some cross sections
in Figure \ref{flugzeug_t}.
 Let us consider more
closely the jammed traffic patterns caused by the temporary bottleneck
associated with the incident. 
While 
the fluctuations are irregular at detector D22 
(about 1.2 km upstream of the aforementioned incident), 
they grow to three stop-and-go waves
further upstream, cf. the arrows
in Fig. \ref{flugzeugV_t}. However, due to measurement errors
and other errors in determining the density \cite{Helb-Opus},
the time series of the density is
very noisy, especially that of cross section D20. 
The adaptive smoothing method suppresses this noise
without suppressing the structure of the pattern.

To check, whether there are significant patterns in the data of detector D20
at all,
we plot the velocity data at D20 as well (Fig. \ref{flugzeugV_t}). 
In contrast to the density, the velocity is measured directly resulting
in less noise. One clearly sees three dips
corresponding to three stop-and-go waves which confirms the results of
the {\em ``adaptive smoothing method''.}

\begin{figure}[htbp] 
\begin{center}
    \includegraphics[width=100\unitlength]{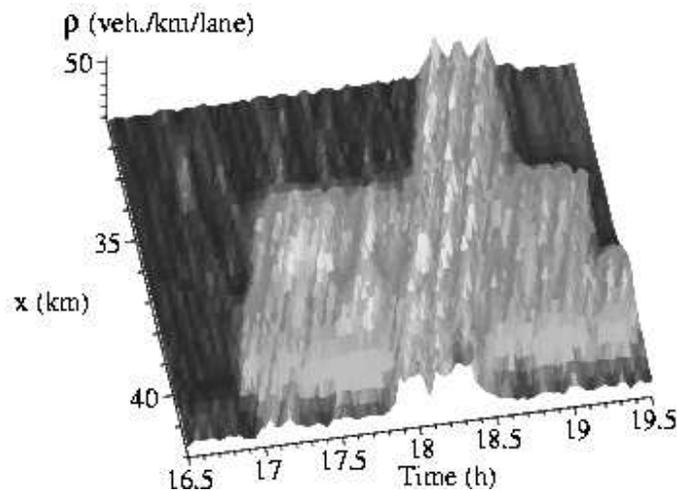}

\end{center}
\caption[]{\label{flugzeug_incomplete}Spatio-temporal 
density  $\rho(x,t)$ as in Figure \protect\ref{flugzeug},
but using only the data of cross sections
D16, D18, D20, and D22 as input (cf. Figure \protect\ref{flugzeug_t}).
}
\end{figure}

As a further check that the structures shown in Fig. \ref{flugzeug}(c)
are not artefacts of the
data processing, we tested the method for a reduced data set.
Figure \ref{flugzeug_incomplete} shows the result using only the
even-numbered detectors as input, whose time series are plotted
in Fig. \ref{flugzeug_t}. 
The main structures remain nearly
unchanged, 
particularly, the three stop-and-go waves propagating through the whole
section.
Notice that  the whole plot is based on data of
only four cross sections, and that the data are  extrapolated
in the upstream direction over a distance of nearly 2 km.

\begin{figure}[htbp] 
\begin{center}
\includegraphics[width=100\unitlength]{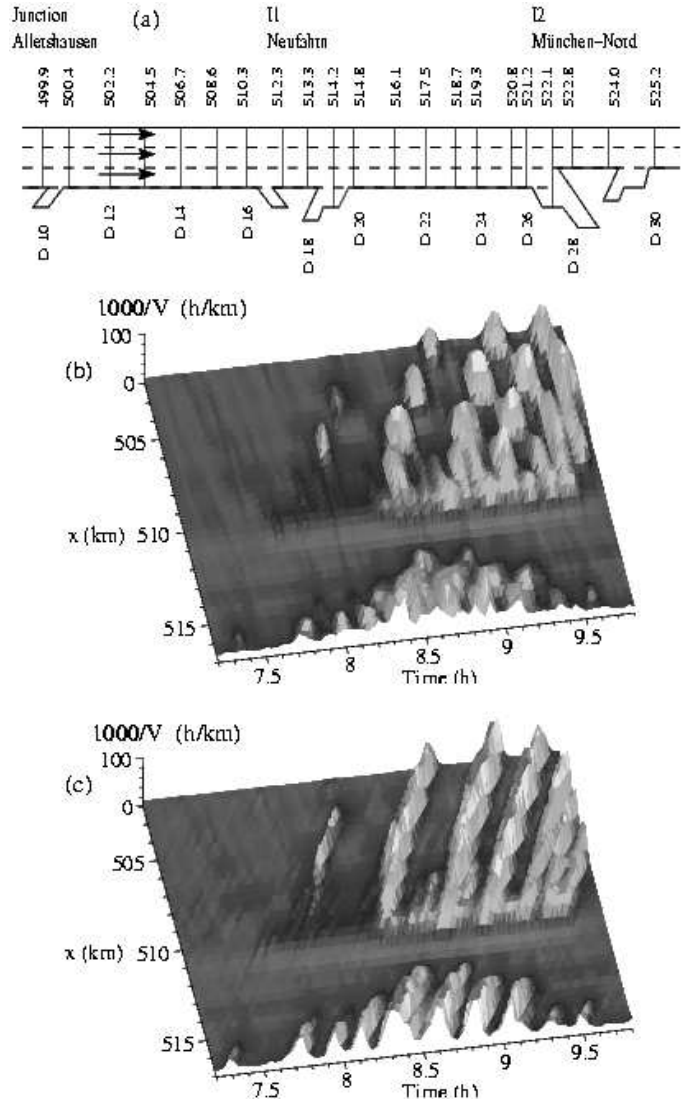}
\end{center}
\caption[]{\label{A9South}Two regions of stop-and-go waves
on the German freeway A9-South near Munich occurring upstream of the 
freeway intersections I1
and I2.
 (a) Sketch of the freeway. (b) 
Plot of the inverse  $1/V(x,t)$ of the spatio-temporal velocity $V(x,t)$, 
using conventional smoothing.
(c) Plot of the same data using
the adaptive smoothing method.
}
\end{figure}

We now apply the processing method to 
data from a section of the freeway  A9-South near Munich.
There are two major intersections I1 and I2 with other 
freeways. Virtually every weekday, traffic broke down to 
oscillatory congested traffic upstream of each of these intersections.
(For details, see Ref.~\cite{Opus}).

Figure \ref{A9South} (b) shows the inverse of the velocity, $1/V(x,t)$ 
for a typical congested situation. When plotting $1/V(x,t)$, the structures of
congested traffic comes out more clearly than
for the density.
The method resolves small density clusters in the region
 508 km $\le x\le 510$~km between about 8:30~am
and 9:30~am which disappear around $x=508$ km.
These structures cannot be identified with the 
conventional  smoothing method (see Fig. \ref{A9South}(c)),
which just shows a hilly pattern. 

Figure \ref{A9South}(b) is an example for the spatial coexistence of
homogeneous congested traffic (a short stretch 
around $x=510$~km), oscillating congested
traffic (around $x=509$~km), and stop-and-go waves (around $x=507$ km).
This may be a three-dimensional illustration of the so-called ``pinch effect''
\cite{KernerPinch,Koshi}, but we do not observe a merging of narrow clusters 
to form wide moving jams. The narrow structures of short wavelengths rather
disappear. This may be an artefact of our smoothing method, so that video data
are required to get a more detailed picture. Simulations of this spatial coexistence, however,
indicate that narrow clusters rather disappear than merge, namely when they do not exceed
the critical amplitude in an area of metastable traffic \cite{PINCH}.

Notice that the bottleneck causing this congestion is located at about
$x=510$ km implying that it is caused by weaving traffic and slowing down at
an off-ramp, not by an on-ramp
as is most often the case on German highways.

\begin{figure}[htbp] 
\begin{center}
    \includegraphics[width=85\unitlength]{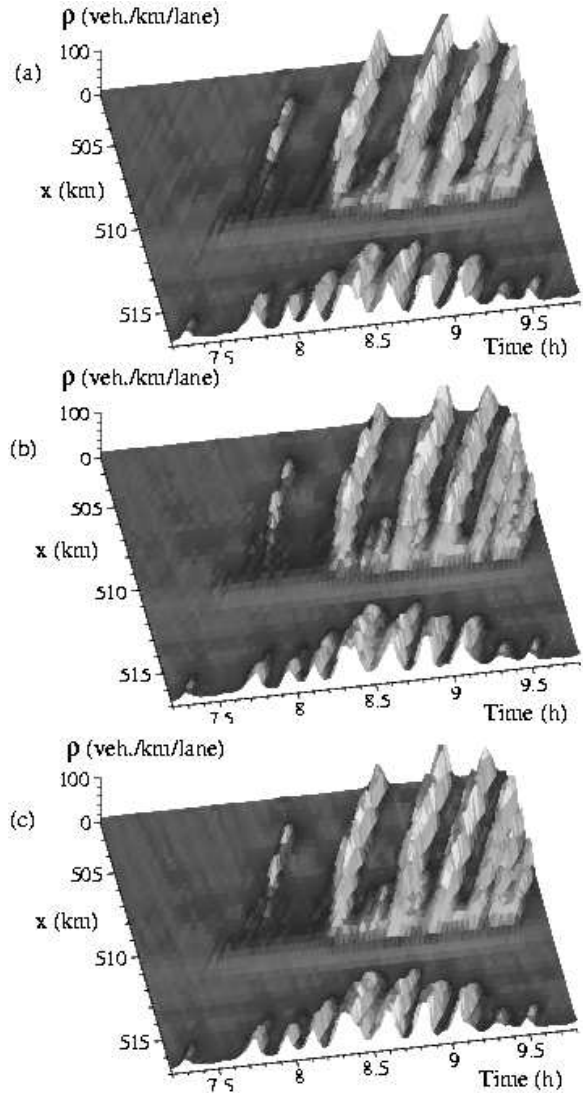} \\[0mm]
  \end{center}
\caption[]{\label{A9South_vg}Same data and same processing method
as in Figure \protect\ref{A9South}(c), but with (a) a propagation velocity
of $c\sub{cong}=-12$~km/h instead of -15~km/h
(b) $c\sub{cong}=-20$~km/h,
(c) with the crossover parameter $V_c=40$ km/h instead of 60 km/h.
}
\end{figure}

Now we demonstrate that the method is robust with respect to reasonable
parameter changes.
Figure \ref{A9South_vg}(a) shows the result for an assumed propagation 
velocity of $c\sub{cong}=-12$~km/h instead of -15~km/h.
In plot (b), we assumed $c\sub{cong}=-20$~km/h instead.
In both cases, the results are systematically better
than with the isotropic procedure. Since the propagation velocity of
perturbations in congested traffic is always about
$c\sub{cong}=-15$ km \cite{Kerner-Rehb96a,Sollacher-Status99,CasMa99,MikaKrYu69}, 
one can take this value as a global setting. 

Due to the
comparatively high magnitude of the propagation velocity $c\sub{free}$,
the related time shifts in the transformation 
\refkl{transform} for free traffic are small. Therefore, the
method is insensitive to changes with respect to this parameter. 
It turned out that taking the isotropic limit
$c\sub{free}\to\infty$ nearly gives the same results.

The crossover parameters $V_c$ and $\Delta V$ can be varied within
reasonable limits as well.
As an example, Fig. \ref{A9South_vg}(c) shows the result after changing
the crossover velocity from $V_c=60$ km/h to
$V_c=40$ km/h.

\begin{figure}[htbp] 
\begin{center}
    \includegraphics[width=100\unitlength]{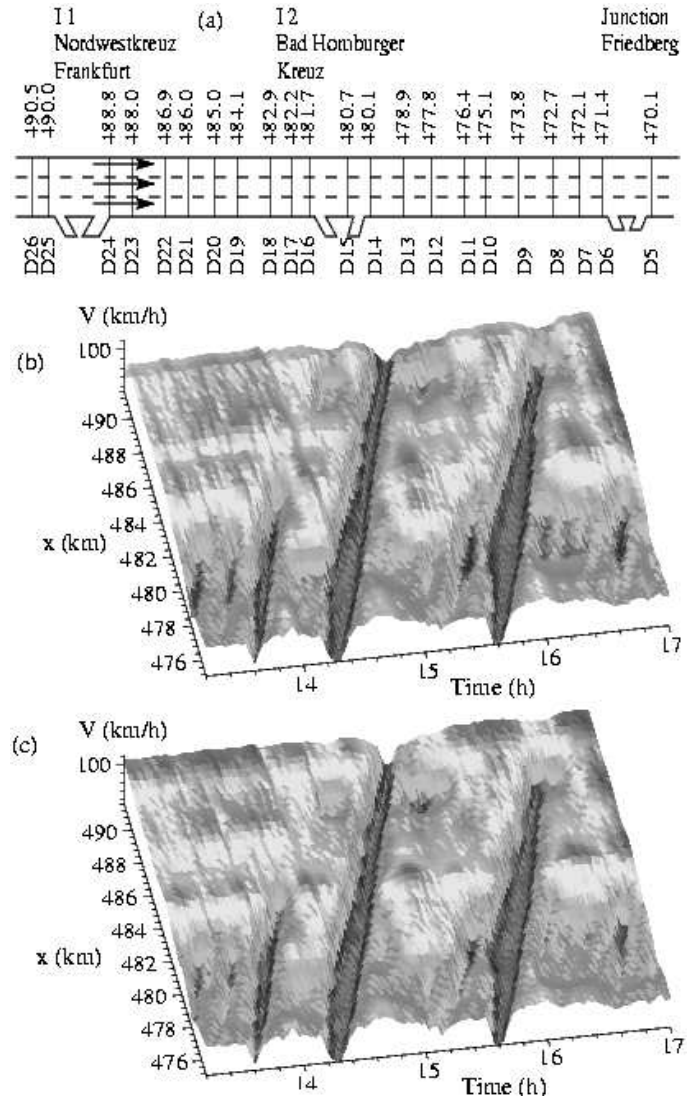} \\[0mm]
\end{center}
\caption[]{\label{A5North}Freeway A5-North from Frankfurt to Kassel
during the evening rush hour
on August 7, 1998.
(a) Sketch of the freeway. (b) Data reconstruction of the velocity 
using all detectors D10--D26. (c) Reconstruction using a reduced data
set of only 6 of all 17 detectors used in (b).
}
\end{figure}
Finally, we applied the method to isolated moving localized clusters
and pinned localized clusters observed on the A5-North near Frankfurt.
This freeway is particularly well equipped 
with detectors, so that the typical length scales of the observed
structures are always larger than the distance between neighbouring
detectors. However, we obtained nearly the same results when
using only the even-numbered detectors  or only
the odd-numbered detectors
as input. Figure \ref{A5North} shows that even a further reduction
of the information to only 35\% of the detectors yields good results.

\section{Summary and Outlook}

We have proposed a new {\em ``adaptive smoothing method''} for the three-dimensional
visualization of spatio-temporal traffic patterns, which takes into account the
characteristic propagation velocities observed in free and congested traffic. The method
is robust with respect to variations of its parameters, so that it can be applied to new
freeway sections without calibration. In principle, it would be possible to determine
the parameters (such as the propagation velocity of perturbations) locally (e.g., by means
of a correlation analysis). However, the results would look less smooth and regular, as the small
number of data to determine the local parameters would be associated with considerable
statistical errors. Consequently, a large part of the variations in the local parameters would
{\em not} reflect {\em systematic} variations of the parameters. So, both the local and global
parameter calibration may produce artefacts, but it is advantageous to use
global parameter settings. Moreover, there is empirical support for surprisingly constant  
propagation velocities $c_{\rm free}$ and $c_{\rm cong}$ 
of perturbations in free and congested traffic. The parameters $V_c$ and $\Delta V$
are related to the transition from free to congested traffic and, therefore, can also be
well determined.  The spatial and temporal smoothing parameters $\sigma$ and $\tau$
can be specified according to the respective requirements. Suitable parameters allow 
a good representation of traffic patterns even when the distances between successive
detectors are about 3 kilometers.
\par
We point out that the suggested {\em ``adaptive smoothing method''} itself can be 
applied to calibrate the characteristic propagation velocities $c_{\rm free}$ and $c_{\rm cong}$. 
For this purpose, the ranges $\sigma$ and $\tau$ of spatial and temporal smoothing 
are chosen small. The optimal propagation velocities minimize the offsets in the propagation
patterns.
\par
The aim of the  {\em ``adaptive smoothing method''} is to reconstruct the spatio-temporal
traffic data from incomplete information as good as possible to allow a better understanding
of the complex traffic dynamics. Potential applications are, for example, traffic state visualization,
incident detection, or the experimental verification of traffic models. Our method
could be further improved by taking into account information about the traffic dynamics
such as the continuity equation or a suitable equation for the average vehicle velocity as a
function of space and time. It could, then, be used to determine short-term traffic forecasts.

\subsection*{Acknowledgments}
The authors would like to thank the German Research Foundation (DFG) 
for financial support through the grant He 2789/2-1. They are also grateful to the
 {\it Autobahndirektion S\"udbayern},
and the {\it Hessisches Landesamt f\"ur Stra{\ss}en und Verkehrswesen}
for providing the traffic data.

\end{document}